\documentclass{jfm}
\usepackage[T1]{fontenc}
\usepackage[latin9]{inputenc}
\usepackage{verbatim}
\usepackage{soul}
\usepackage{xcolor}
\usepackage{graphicx}
\usepackage{amssymb}
\usepackage{amsmath}
\usepackage{esint}
\usepackage{units}
\usepackage[authoryear]{natbib}
\PassOptionsToPackage{normalem}{ulem}
\usepackage{ulem}

\makeatletter
\providecommand{\tabularnewline}{\\}
\providecolor{lyxadded}{rgb}{0,0,1}
\providecolor{lyxdeleted}{rgb}{1,0,0}

\def\vec#1{\mbox{\boldmath $\mathit{#1}$}}
\def\i{\mathrm{i}}

\makeatother

\begin{document}

\title[The effect of finite-conductvity Hartmann walls on the stability of Hunt's flow]
{The effect of finite-conductvity Hartmann walls on the linear stability of Hunt's flow}

\author[T. Arlt, J. Priede, L. Bühler]{Thomas Arlt\aff{1}, J\={a}nis Priede\aff{2} \and~Leo Bühler\aff{1}}
\affiliation{
\aff{1}Institut für Kern- und Energietechnik,\\
Karlsruhe Institute of Technology, Germay\\
\aff{2}Applied Mathematics Research Centre,\\
Coventry University, UK}


\maketitle
\begin{abstract}
\global\long\def\Ha{\mathit{Ha}}
\global\long\def\RE{\mathit{Re}}
We analyse numerically the linear stability of the fully developed
liquid metal flow in a square duct with insulating side walls and
thin electrically conducting horizontal walls with the wall conductance
ratio $c=0.01\cdots1$ subject to a vertical magnetic field with the
Hartmann numbers up to $\Ha=10^{4}.$ In a sufficiently strong magnetic
field, the flow consists of two jets at the side walls walls and a
near-stagnant core with the relative velocity $\sim(c\Ha)^{-1}$.
We find that for $\Ha\gtrsim300,$ the effect of wall conductivity
on the stability of the flow is mainly determined by the effective
Hartmann wall conductance ratio $c\Ha.$ For $c\ll1$, the increase
of the magnetic field or that of the wall conductivity has a destabilizing
effect on the flow. Maximal destabilization of the flow occurs at
$\Ha\approx30/c$. In a stronger magnetic field with $c\Ha\gtrsim30$,
the destabilizing effect vanishes and the asymptotic results of Priede
et al. {[}\emph{J. Fluid Mech.} \textbf{649}, 115, 2010{]} for the
ideal Hunt's flow with perfectly conducting Hartmann walls are recovered. 
\end{abstract}

\section{Introduction}

Some tokamak-type nuclear fusion reactors, which are expected to provide
virtually unlimited amount of safe energy in the future, contain blankets
made of rectangular ducts in which liquid metal flows in a high, transverse
magnetic field between $\unit[5]{T}$ and $\unit[10]{T}$ \citep{Buehler2007}.
These blankets are designed to cool the plasma chamber, to breed and
to remove tritium as well as to protect the superconducting magnetic
field coils from the neutron radiation emitted by the fusion plasma.
The transfer properties of the magnetohydrodynamic flows depend strongly
on their stability. On the one hand, magnetohydrodynamic instabilities
and the associated turbulent mixing can enhance the transport of heat
and mass, which is beneficial for the cooling and removal of tritium.
On the other hand, it can also enhance the transport of momentum,
which has an adverse effect on the hydrodynamic resistance of the
duct \citep{Zikanov2014}. 

Linear stability of MHD flows strongly varies with the electrical
conductivity of the duct walls. In the duct with perfectly conducting
walls, where the flow has weak jets along the walls parallel to the
magnetic field \citep{Uflyand1961,Chang1961}, the critical Reynolds
number based on the maximum flow velocity increases asymptotically
as $\RE_{c}\sim642\Ha^{1/2},$ where the Hartmann number $\Ha$ defines
the strength of the applied magnetic field \citep{Priede2012}. In
the duct made of thin conducting walls, where strong side-wall jets
carry a significant fraction of the volume flux \citep{Walker1981},
the flow becomes unstable at a substantially lower maximal velocity
corresponding to $\RE_{c}\sim110\Ha^{1/2}$ \citep{PriedeArltBuehler2015}.
Even more unstable is the so-called Hunt's flow \citep{Hunt1965},
which develops when the walls parallel to the magnetic field are insulating
whereas the perpendicular walls, often referred to as the Hartmann
walls, are perfectly conducting. In this case the side-wall jets carry
dominant part of the volume flux, and the asymptotic instability threshold
drops to $\RE_{c}\sim90\Ha^{1/2}$ \citep{Priede2010}.

Hunt's flow is conceptually simple but like the flow in perfectly
conducting duct it is rather far from reality where the walls usually
have a finite electrical conductivity. Therefore it is of practical
importance to consider the effect of finite electrical conductivity
of Hartmann walls on the experimentally viable Hunt's flow. This is
the main focus of the present study which is concerned with linear
stability analysis of the imperfect Hunt's flow with thin finite conductivity
Hartmann walls.

The paper is organised as follows. The problem is formulated in $\S$\ref{sec:prob}.
Numerical results for a square duct are presented in $\S$\ref{sec:res}.
The paper is concluded with a summary and discussion of results in
$\S$\ref{sec:sum}.

\begin{figure}
\begin{centering}
\includegraphics[width=0.4\textwidth]{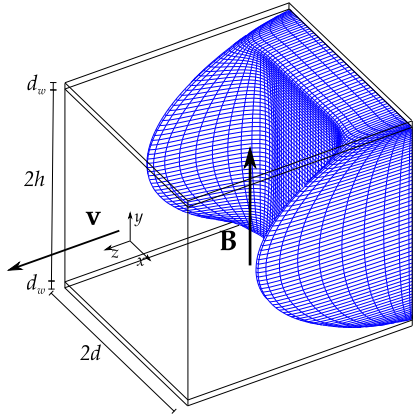}
\par\end{centering}
\caption{\label{fig:sketch}The base flow profile in a square duct with insulating
side walls and thin conducting horizontal walls with the wall conductance
ratio $c=0.1$ subject to a vertical magnetic field with $\textit{\ensuremath{\protect\Ha}}=100.$ }
\end{figure}

\section{\label{sec:prob}Formulation of the problem}

Consider the flow of an incompressible viscous electrically conducting
liquid in a duct with half width $d$ and half height $h$ inside
a transverse homogeneous magnetic field $\vec{B}$. The point of origin
is at the centre of the duct with axis orientation as shown in figure
\ref{fig:sketch}. The liquid flow is governed by the Navier-Stokes
equation
\begin{equation}
\partial_{t}\vec{v}+(\vec{v}\cdot\vec{\nabla})\vec{v}=-\rho^{-1}\vec{\nabla}p+\nu\vec{\nabla}^{2}\vec{v}+\rho^{-1}\vec{f},\label{eq:NS}
\end{equation}
with the electromagnetic body force $\vec{f}=\vec{j}\times\vec{B}$
involving the induced electric current $\vec{j},$ which is governed
by the Ohm's law for a moving medium 
\begin{equation}
\vec{j}=\sigma(\vec{E}+\vec{v}\times\vec{B}).\label{eq:Ohm}
\end{equation}
The flow is assumed to be sufficiently slow for the induced magnetic
field to be negligible relative to the imposed one. This corresponds
to the so-called inductionless approximation which holds for small
magnetic Reynolds numbers \global\long\def\Rm{\mathit{Rm}}
$\Rm=\mu_{0}\sigma v_{0}d\ll1,$ where $\mu_{0}$ is the permeability
of free space and $v_{0}$ is a characteristic velocity of the flow.
In addition, we assume the characteristic time of velocity variation
to be much longer than the magnetic diffusion time $\tau_{m}=\mu_{0}\sigma d^{2}.$
This is known in MHD as the quasi-stationary approximation \citep{Roberts1967},
which leads to $\vec{E}=-\vec{\nabla}\phi,$ where $\phi$ is the
electrostatic potential. 

Velocity and current satisfy mass and charge conservation $\vec{\nabla}\cdot\vec{v}=0,\vec{\nabla}\cdot\vec{j}=0.$
Applying the latter to Ohm's law (\ref{eq:Ohm}) and using the inductionless
approximation, we obtain

\begin{equation}
\vec{\nabla}^{2}\phi=\vec{B}\cdot\vec{\omega},\label{eq:phi}
\end{equation}
where $\vec{\omega}=\vec{\nabla}\times\vec{v}$ is the vorticity.
At the duct walls $S$, the normal $(n)$ and tangential $(\tau)$
velocity components satisfy the impermeability and no-slip boundary
conditions $\left.v_{n}\right\vert _{s}=0$ and $\left.v_{\tau}\right\vert _{s}=0.$
Charge conservation applied to the thin wall leads to the following
boundary condition 
\begin{equation}
\left.\partial_{n}\phi-dc\vec{\nabla}_{\tau}^{2}\phi\right\vert _{s}=0,\label{bc:phi}
\end{equation}
 where $c=\sigma_{w}d_{w}/(\sigma d)$ is the wall conductance ratio
\citep{Walker1981}. At the non-conducting side walls with $c=0$
we have $\partial_{n}\phi|_{s}=0$.

The problem admits a rectilinear base flow along the duct with $\bar{\vec{v}}=(0,0,\bar{w}(x,y)),$
which is computed numerically and then analysed for the linear stability
using a vector streamfunction-vorticity formulation introduced by
\citealt{Priede2010} which briefly outlined in the Appendix. 

\section{\label{sec:res}Results}

\begin{figure}
\begin{centering}
\includegraphics[width=0.5\textwidth]{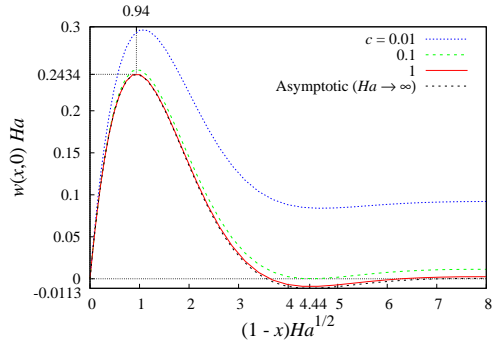} 
\par\end{centering}
\caption{\label{fig:bflow} Horizontal base flow velocity profiles at $y=0$
in the vicinity of the side-wall in stretched coordinate $(1-x)\protect\Ha^{1/2}$
for various Hartmann wall conductance ratios $c$ and $\protect\Ha=10^{3}.$ }
\end{figure}

Let us first consider the principal characteristics of the base flow,
which will be useful for interpreting its stability later. Although
rectangular duct with insulating side walls and thin conducting Hartmann
walls admits an analytical Fourier series solution, \citep{Hunt1965}
it is more efficient to compute the base flow numerically \citep{Priede2010}.
On the other hand, the important properties of the base flow can be
deduced from the general asymptotic solution derived by \citet{PriedeArltBuehler2015}
for arbitrary Hartmann and side wall conductane ratios $c_{n}$ and
$c_{\tau}.$ Below we provide the relevant results which follow from
the general solution for the case of Hunt's flow with thin Hartmann
walls $(c_{n}=c>0)$ and insulating side walls $(c_{\tau}=0).$ According
to the asymptotic solution, the conductivity of Hartmann walls affects
primarily the flow in the core region of the duct. In a sufficiently
strong magnetic satisfying $c\Ha\gg1,$ the core velocity scales as
\begin{equation}
\bar{w}_{\infty}\sim-(1+c^{-1})\Ha^{-2}\bar{P}.\label{eq:winf}
\end{equation}
Velocity distribution at the side walls can written as 
\begin{equation}
\bar{w}_{0}(\tilde{x},y)=\bar{w}_{\infty}[1+\sum_{k=0}^{\infty}e^{-\lambda\tilde{x}}B\left[C\sin(\lambda\tilde{x})+\cos(\lambda\tilde{x})\right]\cos(\kappa y)],\label{eq:w0}
\end{equation}
where $\tilde{x}=\Ha^{1/2}(A\pm x)$ is a stretched side-layer coordinate
and $A=h/d$ is the aspect ratio; $\kappa=\pi(k+1/2),$ $\lambda=\sqrt{\kappa/2},$
$B=(-1)^{k+1}2/\kappa$ and $C=-\Ha^{-1}\bar{P}/(\bar{w}_{\infty}\kappa)$
are coefficients which depend on the summation index $k$. For $c\Ha\gg1$,
the conductivity of Hartmann walls has virtually no effect on the
velocity distribution (\ref{eq:w0}), which reduces to that of the
ideal Hunt's flow with the maximal jet velocity 
\begin{equation}
\bar{w}_{\max}\sim-0.2434\Ha^{-1}\bar{P},\label{eq:wmax}
\end{equation}
which is located at the distance $\delta\sim0.94\Ha^{-1/2}$ from
the side wall (see Fig. \ref{fig:bflow}). As seen, the velocity of
jets is much higher than that of the core if $\Ha\gg1+c^{-1}.$ For
poorly conducting Hartmann walls with $c\ll1$, this is condition
reduces to $c\Ha\gg1.$ It is the same condition that underlies (\ref{eq:winf})
and means that the Hartmann walls are well conducting relative to
the adjacent Hartmann layers. In this the case, the velocity of core
flow scales as $\sim(c\Ha)^{-1}$ relative to that of side-wall jets.
It means that the effect of the core flow and, consequently, that
of the conductivity of Hartmann walls on the stability of Hunt's flow
is expected to vanish when the jet velocity is used to parametrize
the problem.

\begin{figure}
\begin{centering}
\includegraphics[width=0.5\textwidth]{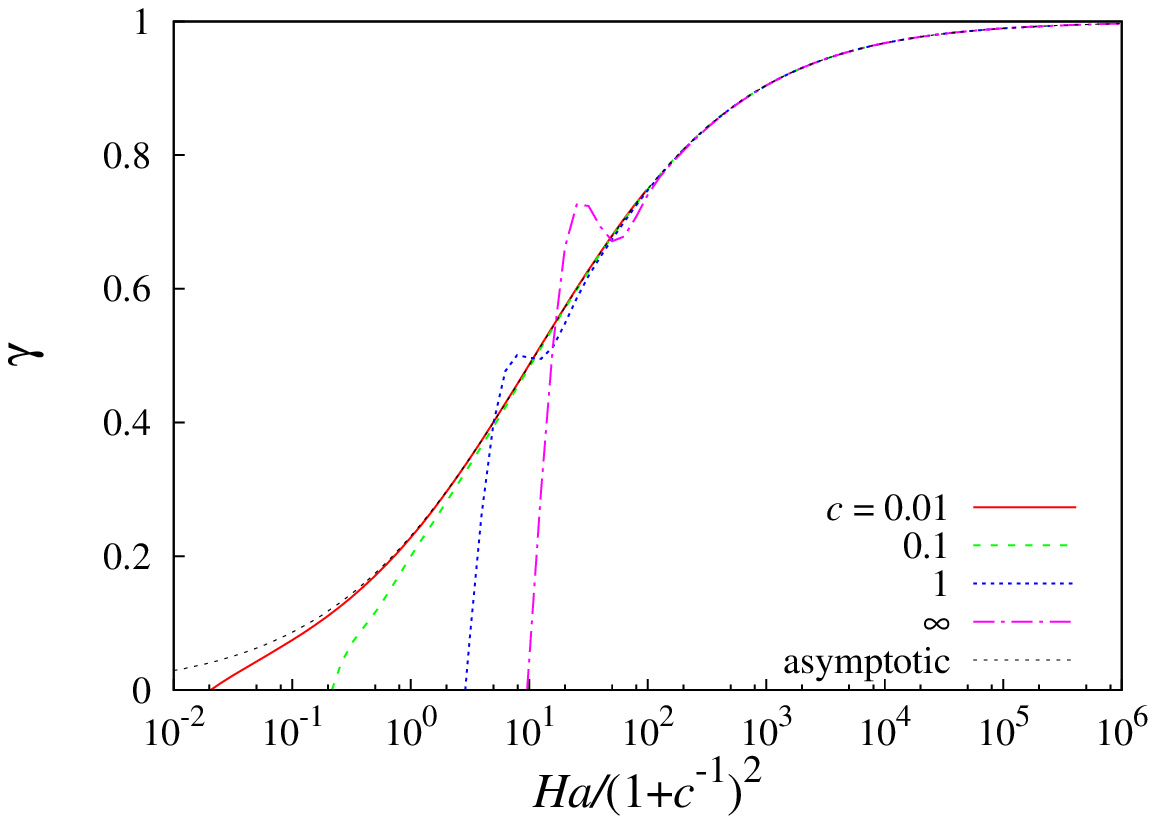}\put(-180,135){(\textit{a})}\includegraphics[width=0.5\textwidth]{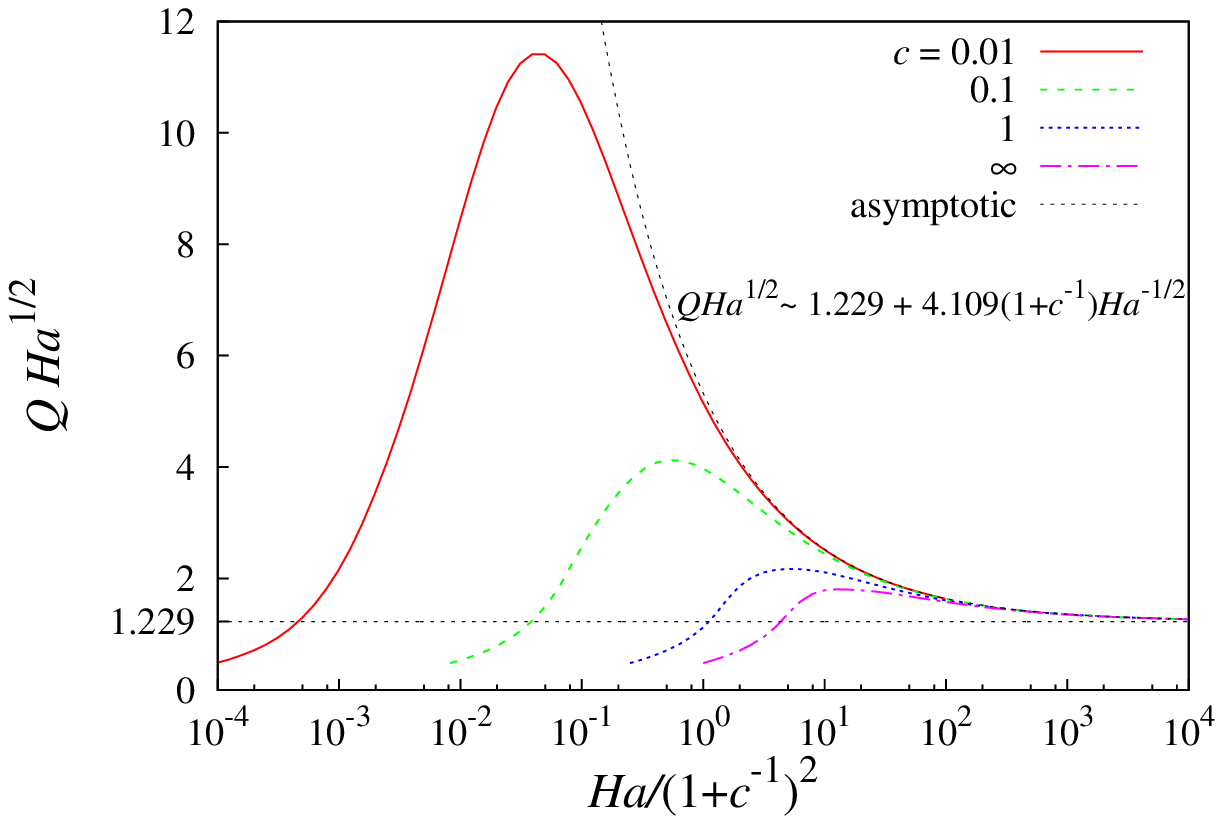}\put(-180,135){(\textit{b})}
\par\end{centering}
\caption{\label{fig:flow_rate}The volume flux fraction $\gamma$ carried by
the side-wall jets for different wall conductance ratios $c$ (a)
and the rescaled total volume flux $Q\protect\Ha^{1/2}$ (b) versus
the modified Hartmann number $\protect\Ha/(1+c^{-1})^{2}$ for the
base flow normalized with the maximal jet velocity; $c=\infty$ corresponds
to the ideal Hunt's flow with perfectly conducting Hartmann walls. }
\end{figure}

The volume flux carried by the side-wall jets in a quarter duct cross-section
is
\[
q=\Ha^{-1/2}\int_{0}^{\infty}\int_{0}^{1}\left(\bar{w}_{0}(\tilde{x},y)-\bar{w}_{\infty}\right)\mathrm{d}y\thinspace\mathrm{d}\tilde{x}\sim\alpha\Ha^{-3/2}\bar{P},
\]
 where $\alpha=(16-\sqrt{2})\pi^{-7/2}\zeta(\frac{7}{2})\approx0.299$
and $\zeta(x)$ is the Riemann zeta function \citep{Abramowitz1964}.
The respective fraction of the volume flux carried by the side-wall
jets is
\[
\gamma\sim\frac{q}{q+\bar{w}_{\infty}A}=\left(1+A\alpha^{-1}(1+c^{-1})\Ha^{-1/2}\right)^{-1}.
\]
The expression above is confirmed by the numerical results plotted
in figure \ref{fig:flow_rate}(a), where the curves for different
wall conductance ratios are seen to collapse to this asymptotic solution
when plotted against the modified Hartmann number $\Ha/(1+c^{-1})^{2}.$
It implies that a strong magnetic field with $\Ha\gg c^{-2}$ is required
for the side-wall jets to fully develop, that is to carry the dominant
fraction of the volume flux $\gamma\rightarrow1$ in the non-ideal
Hunt's flow with weakly conducting Hartmann walls $(c\ll1).$ This
is confirmed also by the total volume flux plotted in figure \ref{fig:flow_rate}(b)
for the base flow normalized with the maximal velocity, which we use
as the characteristic velocity in this study. The curves for different
conductance ratios are seen to collapse to the asymptotic solution
\begin{equation}
Q=(q+\bar{w}_{\infty}A)/\bar{w}_{\max}\sim(1.229+4.109(1+c^{-1})\Ha^{-1/2})\Ha^{-1/2}\label{eq:Q}
\end{equation}
when $\Ha/(1+c^{-1})^{2}\gg1.$ This shows that a much stronger magnetic
field is required to attain the asymptotic regime in the non-ideal
Hunt's flow when the volume flux rather than the jet velocity is used
to parametrise the problem. It is due to the much larger area of the
core region, which scales as $\sim\Ha^{1/2}$ relative to that of
the side-wall jets and thus makes the fraction of the volume flux
carried by the core flow $\sim1/(c\Ha^{1/2}).$ Therefore it is advantageous
to use the maximal rather than the mean velocity as a characteristic
parameter. In this way, the high-field asymptotics of critical parameters
can be extracted from the numerical solution at significantly lower
Hartmann numbers. One can use the volume flux plotted in figure \ref{fig:flow_rate}(a)
or the asymptotic expression (\eqref{eq:Q}), if the magnetic field
is sufficiently strong, to convert from our Reynolds number based
on the maximal velocity to that based on the mean velocity.

\begin{figure}
\begin{centering}
\includegraphics[width=0.5\textwidth]{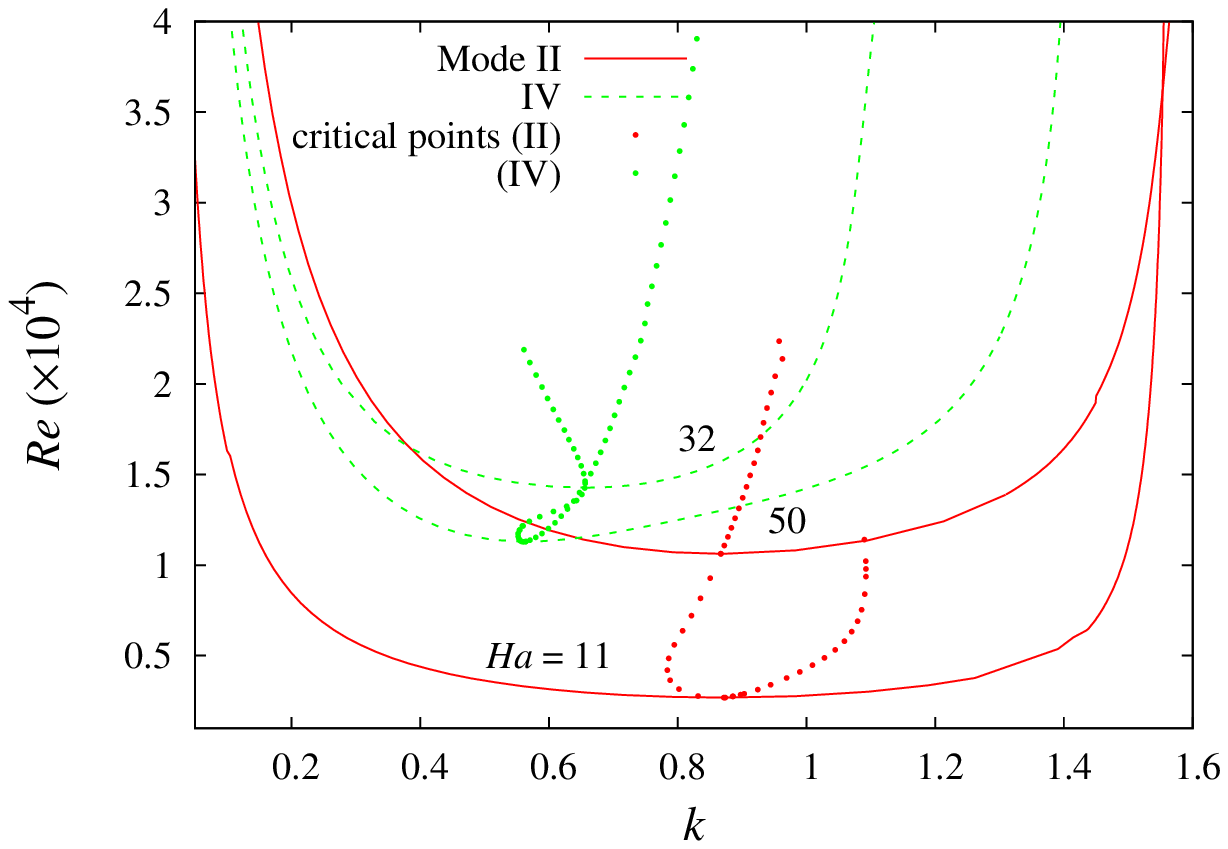}\put(-180,135){(\textit{a})}\includegraphics[width=0.5\textwidth]{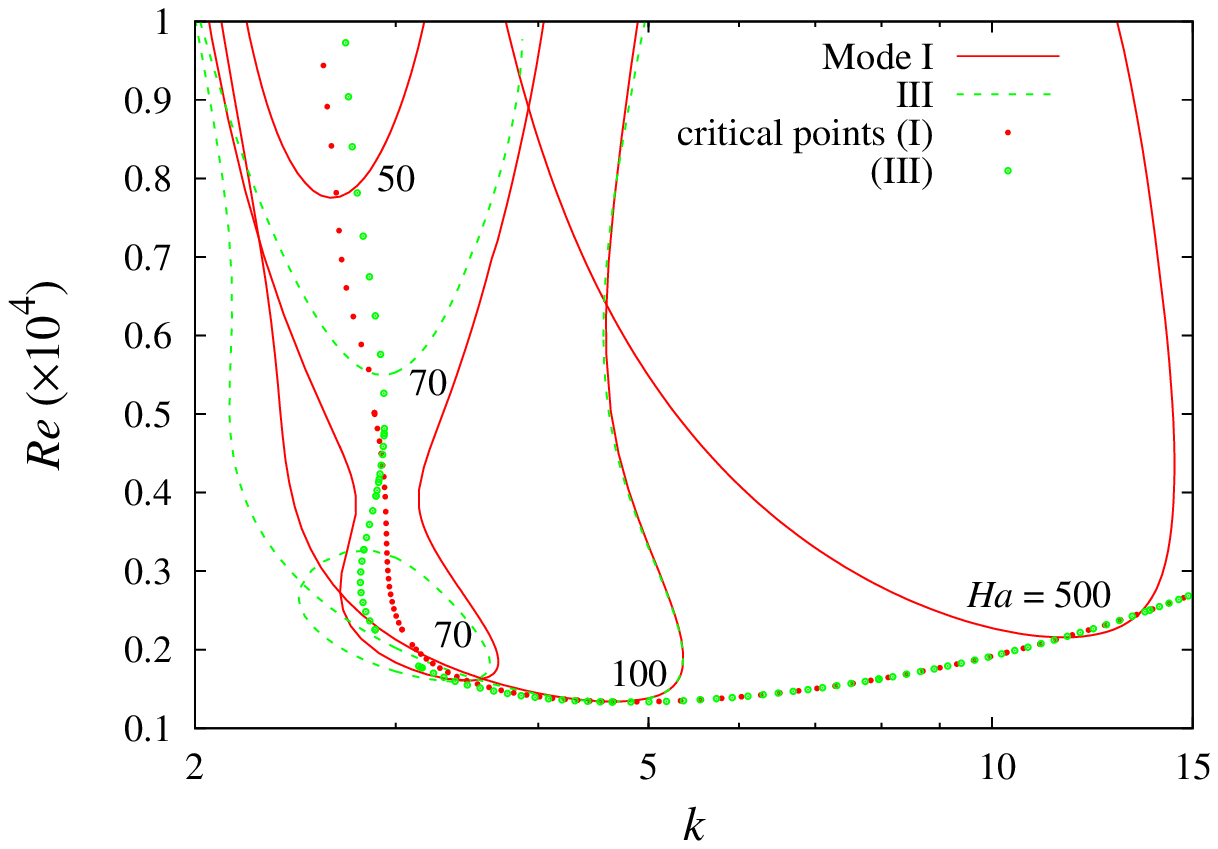}\put(-180,135){(\textit{b})}
\par\end{centering}
\caption{Marginal Reynolds numbers for modes II,IV (a)and I,III (b) depending
on wavenumber $k$ for $c=1$. The dots indicate critical points for
other Hartmann numbers.\label{fig:Re-k}}
\end{figure}

\begin{figure}
\begin{centering}
\includegraphics[width=0.5\textwidth]{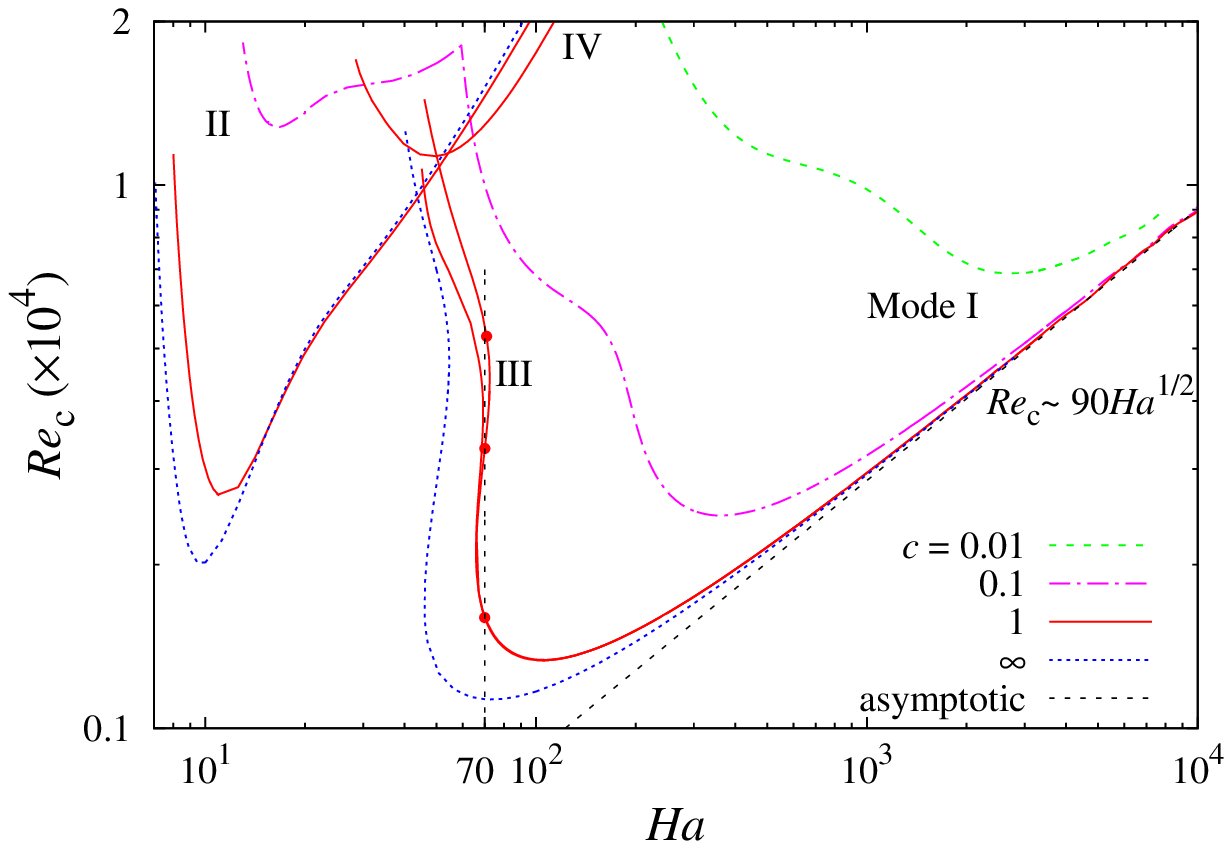}\put(-180,135){(\textit{a})}\includegraphics[width=0.5\textwidth]{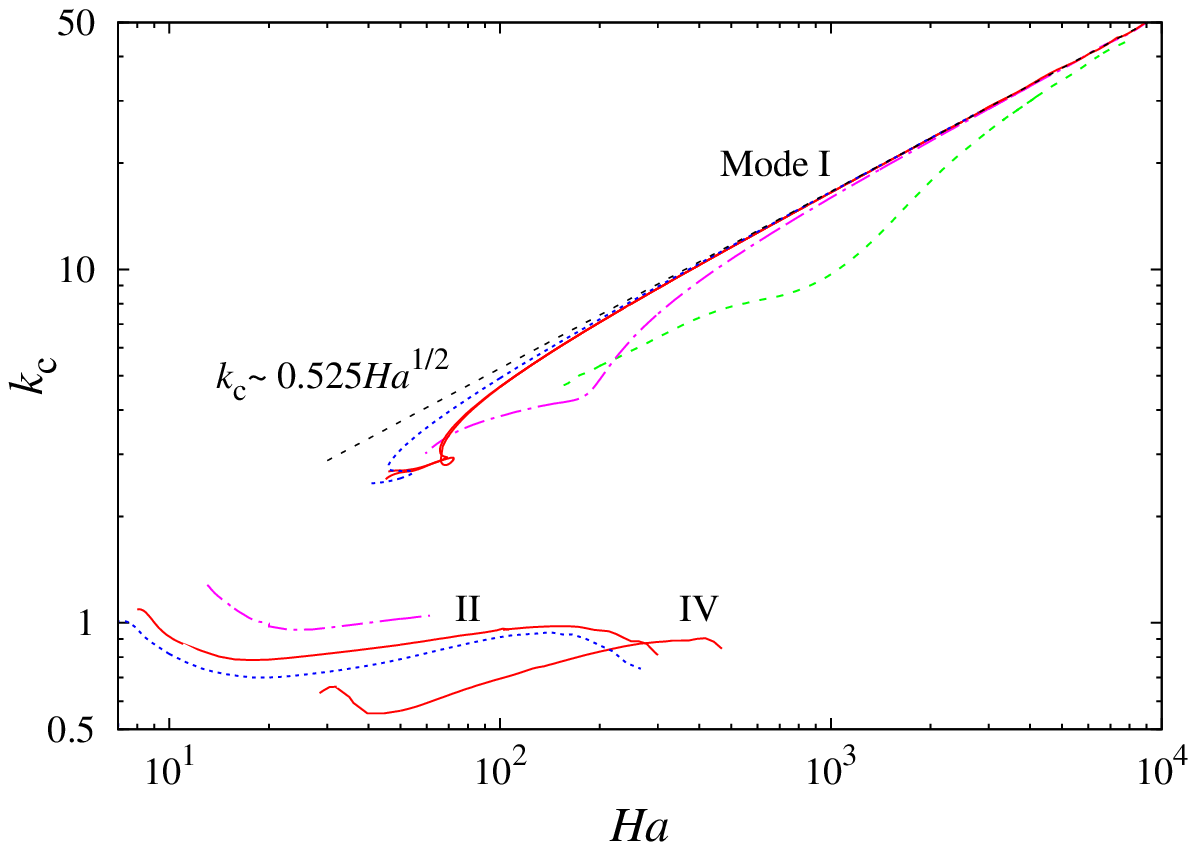}\put(-180,135){(\textit{b})}\\
\includegraphics[width=0.5\textwidth]{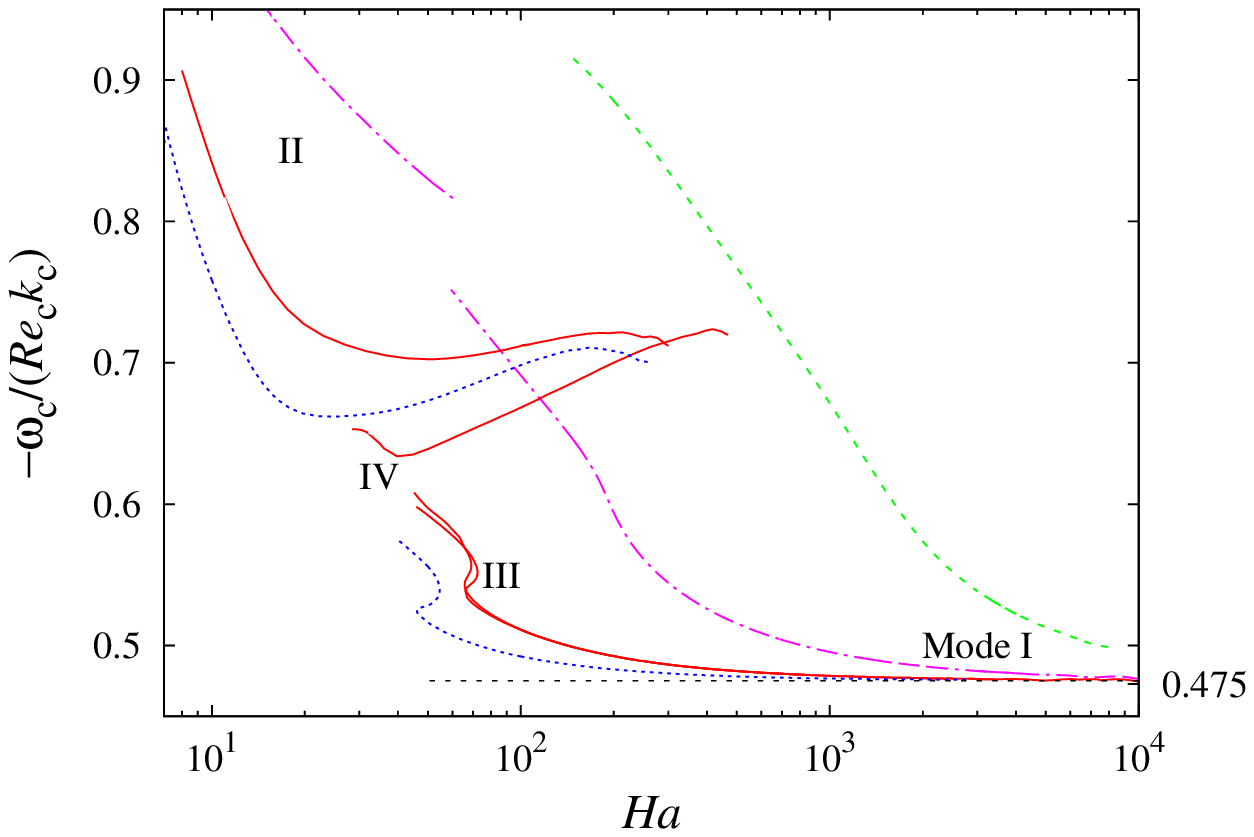}\put(-180,135){(\textit{c})}
\par\end{centering}
\caption{\label{Rec-Ha} Critical Reynolds number (a), wavenumber (b) and the
relative phase velocity (c) against the Hartmann number for different
wall conductance ratios $c$; $c=\infty$ corresponds to the ideal
Hunt's flow considered by \citet{Priede2010}. }
\end{figure}

Now let us turn to the stability of the flow in a square duct $(A=1)$
and start with moderately conducting Hartmann walls with $c=1,$ which
is used in the following unless stated otherwise. The marginal Reynolds
number, at which the growth rate of different instability modes turns
zero are plotted against the wavenumber in figure \ref{fig:Re-k}.
The minimum on each marginal Reynolds number curve defines a critical
wavenumber and the respective Reynolds number $\RE_{c}$ by exceeding
which the flow becomes unstable with respect to the perturbation of
the given symmetry for the specified Hartmann number. Other critical
points for different instability modes and Hartmann numbers are marked
by dots in figure \ref{fig:Re-k} and plotted explicitly against the
Hartmann number in figure \ref{Rec-Ha} with solid lines for $c=1$
and with dashed lines for other wall conductivity ratios. For $c=1,$
figure \ref{Rec-Ha} shows critical parameters for all four instability
modes. For other wall conductance ratios, only the critical parameters
for the most unstable modes, i.e. those with the lowest $\RE_{c}$
for the given $\Ha,$ are plotted. Figure \ref{Rec-Ha} (a) shows
that for $c=1,$ a magnetic field with $\Ha\gtrsim8$ is required
for the flow in square duct to become linearly unstable. The first
instability mode is of type II, which is typical for the MHD duct
flows. Critical Reynolds number for this mode reaches minimum $\RE_{c}\approx12000$
at $\Ha\approx13$ and then starts to increase with the magnetic field.
In a high magnetic field, the increase of critical Reynolds number
is close to $\RE_{c}\sim\Ha.$

At $\Ha\gtrsim25,$ an instability mode of symmetry type IV appears
with a critical Reynolds number significantly higher than that for
the mode II. Critical Reynolds number for this mode attains minimum
$\RE_{c}\approx17000$ at $\Ha\approx50$ and then stays slightly
below $\RE_{c}$ for mode II at larger $\Ha.$ These two instability
modes are efficiently stabilized by a sufficiently strong magnetic
field which leads to $\RE_{c}$ increasing nearly linearly with $\Ha.$
This is due to the anti-symmetric vertical distribution of the $y$-component
of vorticity (see table \ref{tab:mod}), which makes it essentially
non-uniform along the magnetic field and thus subject to a strong
magnetic damping.

Two additional instability modes \textendash{} one of type I and another
of type III \textendash{} emerge at $\Ha\approx50.$ These two modes
have similar critical Reynolds numbers, which are seen in figure \ref{Rec-Ha}(a)
to quickly drop below those for modes II and IV. This makes modes
I and III the most unstable ones in a sufficiently strong magnetic
field. The critical wavenumbers for modes I and III are practically
indistinguishable one from another in figure \ref{Rec-Ha}(b). As
seen in figure \ref{fig:Re-k}(b), at moderate Hartmann numbers $\Ha\approx70$
modes I and III may have intricate neutral stability curves which
of consist of closed contours and disjoint open parts. For $\Ha=70,$
mode III has not one but three critical Reynolds numbers. By exceeding
the lowest $\RE_{c}$ the flow becomes unstable and remains such up
to the second $\RE_{c},$ above which it restabilizes. After that
the flow remains stable up to the third $\RE_{c},$ above which it
turns ultimately unstable. As seen in figure \ref{Rec-Ha}(a), such
a triple stability threshold for modes I/III, which are shown by dots
for mode III at $\Ha=70$, exists only in a relative narrow range
of Hartmann numbers around $\Ha\approx70.$ 

The lowest critical Reynolds number for modes I and III, $\RE_{c\ }\approx1300$
is attained at $\Ha\approx100.$ In a high magnetic field, the critical
Reynolds numbers and wavenumbers for both modes increase asymptotically
as $\RE_{c}\sim k_{c}\sim\Ha^{1/2}$ , which means that the relevant
length scale of instability is determined by the thickness of the
side layers $\delta\sim\Ha^{-1/2}.$ 

The much weaker magnetic damping of modes I/III in comparison to modes
II/IV is due to the symmetric (even) vertical distribution of the
$y$-component of vorticity which makes it relatively uniform along
the magnetic field. The spanwise symmetry, which is opposite for modes
I and III, and determines the relative sense of rotation of vertical
vortices at the opposite side walls, has almost no effect on their
stability in a sufficiently strong magnetic field, where the critical
parameters for both modes become virtually identical.

Now let us consider the stability of the flow at lower wall conductance
ratios. As seen in figure \ref{Rec-Ha}(a), the critical Reynolds
number increases, which means that the flow becomes more stable, when
the wall conductivity is reduced provided that the magnetic field
is not too strong. In a strong magnetic field, this stabilizing effect
vanishes and the curves for different wall conductance ratios collapse
to the asymptotics of the ideal Hunt's flow with $\RE_{c}\sim90\Ha^{1/2},$
$k_{c}\sim0.525\Ha^{1/2}$ and $-\omega_{c}/(\RE_{c}k_{c})\approx0.475$
found by \citet{Priede2010}. 

\begin{figure}
\centering{}\includegraphics[width=0.5\textwidth]{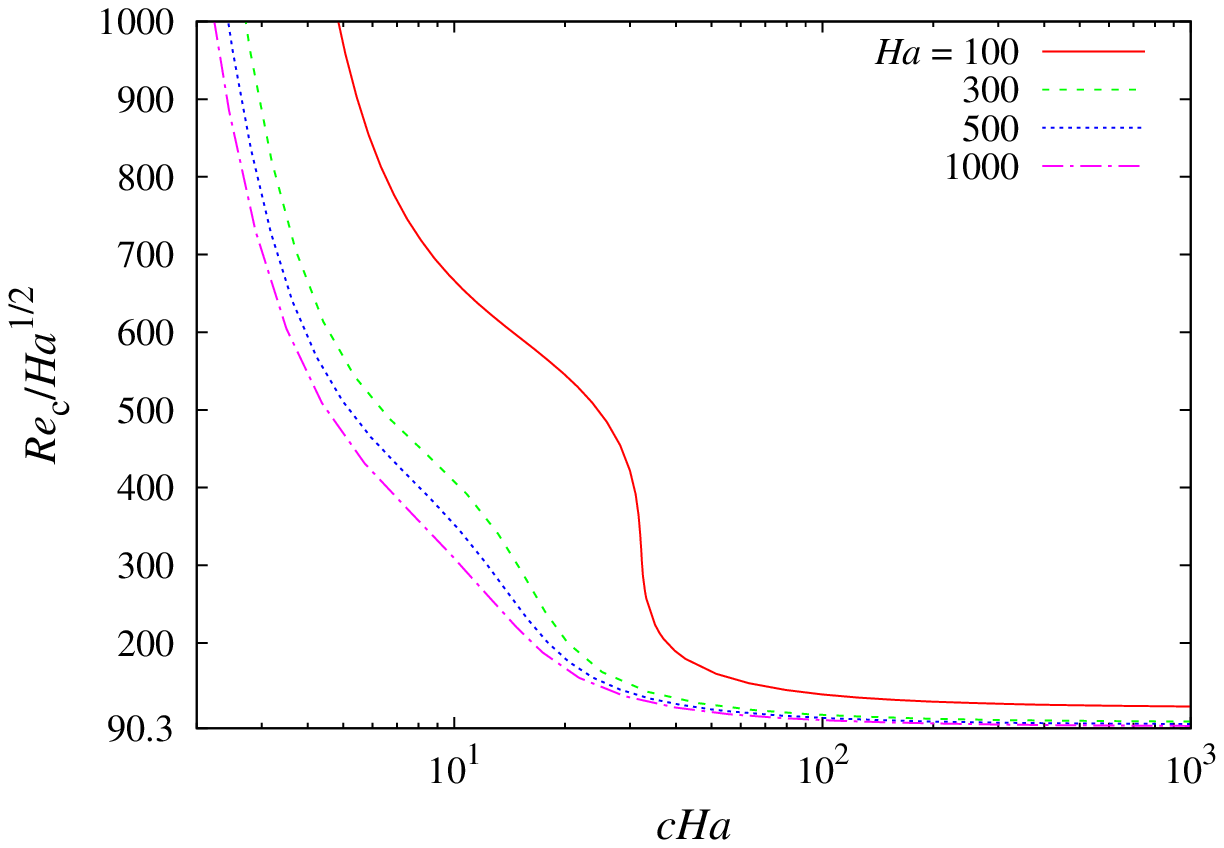}\put(-180,135){(\textit{a})}\includegraphics[width=0.5\textwidth]{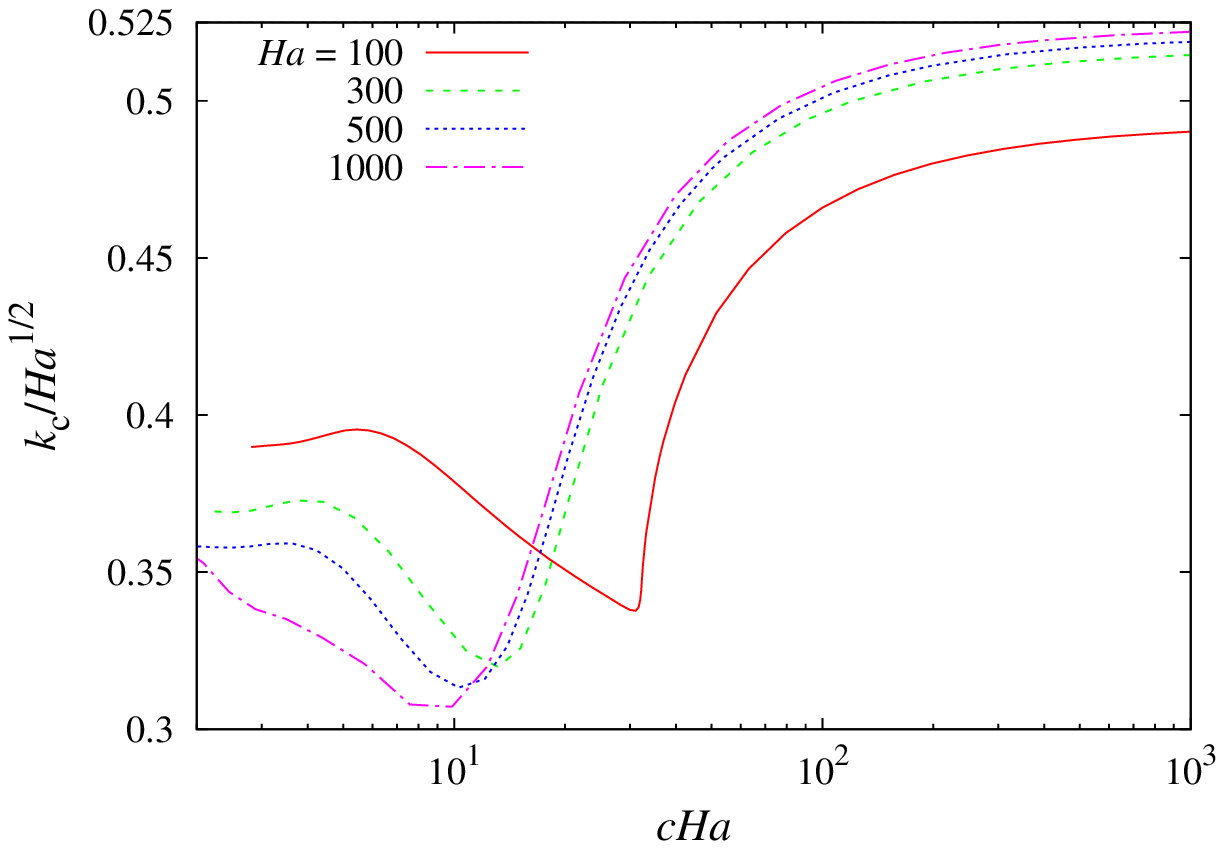}\put(-180,135){(\textit{b})}\caption{\label{fig:Rec-c}Rescaled critical Reynolds number $\protect\RE_{c}/\protect\Ha^{1/2}$
(a) and the wavenumber $k_{c}/\protect\Ha^{1/2}$ (b) against the
rescaled wall conductance ratio $c\protect\Ha$ for different Hartmann
numbers $\protect\Ha.$}
\end{figure}

As discussed at the beginning of this section, $\Ha\gg c^{-1}$ is
required for these asymptotics. This estimate is consistent with the
Hartmann numbers at which the lowest critical Reynolds numbers are
attained for different wall conductance ratios (see table \ref{tab:REc_min}).
A more specific confirmation of this estimate may be seen in figure
\ref{fig:Rec-c}(a), where the rescaled critical Reynolds numbers
$\RE_{c}/\Ha^{1/2}$ collapse for different $\Ha\gtrsim300$ to nearly
the same curve when plotted against the rescaled wall conductance
ratio $c\Ha.$ Also the rescaled critical wavenumber $k_{c}/\Ha^{1/2}$
is seen in figure \ref{fig:Rec-c}(b) to collapse in a similar way
for $c\Ha\gtrsim10.$ The latter parameter combination represents
an effective wall conductance ratio which defines the wall conductance
relative to that of the adjacent Hartmann layer. A more detailed physical
interpretation of $c\Ha$ will be given in the conclusion. Figure
\ref{fig:Rec-c} shows that in a strong magnetic field with $\Ha\gtrsim300,$
the effect of wall conductivity on the stability of flow is mainly
determined by a single parameter, the effective wall conductance ratio
$c\Ha.$ According to table \ref{tab:REc_min}, Hartmann walls with
$c\ll1$ have a stabilizing effect on the flow only up to $c\Ha\gtrsim30.$
At higher effective wall conductance ratios, as seen in figure \ref{fig:Rec-c},
the high-field asymptotics of Hunt's flow are recovered.

\begin{table}
\begin{centering}
\begin{tabular}{ccc}
$c$ & $\RE_{c}$ & $\Ha$\tabularnewline
1 & 1300 & 100\tabularnewline
0.1 & 2500 & 350\tabularnewline
0.01 & 6900 & 2800\tabularnewline
\end{tabular}
\par\end{centering}
\caption{\label{tab:REc_min}The lowest critical Reynolds numbers $\protect\RE_{c}$
and the corresponding Hartmann numbers for different wall conductance
ratios $c$}
\end{table}

\section{Summary and Conclusions\label{sec:sum}}

We have investigated numerically the linear stability of a realistic
Hunt's flow in a square duct with finite conductivity Hartmann walls
and insulating side walls subject to a homogeneous vertical magnetic
field. It was found that in a sufficiently strong magnetic field with
$\Ha\gtrsim300$ the impact of wall conductivity on the stability
of the flow is determined mainly by a single parameter \textendash{}
the effective Hartmann wall conductance ratio $c\Ha$. This parameter
characterizes the wall conductance relative to that of the adjacent
Hartmann layer, which is connected electrically in parallel to the
former. Because the thickness and, thus, also the conductance of Hartmann
layer drops inversely with the applied magnetic field, a finite conductivity
wall becomes relatively well conducting in the sufficiently strong
magnetic field. 

There are two reasons why $c\Ha$ emerges as the relevant stability
parameter of Hunt's flow with finite-conductivity Hartmann walls.
Firstly, it is due to the core region of the base flow, whose velocity
in a sufficiently strong magnetic field scales according to (\ref{eq:winf})
as $\sim(c\Ha)^{-1}$ relative to that of the side-wall jets (\ref{eq:wmax})
when the wall conductance ratio is small $(c\ll1)$. Secondly, $c\Ha$
plays the role of effective wall conductance ratio in the thin-wall
boundary condition (\ref{bc:phi}) and replaces $c$ when the side-wall
jet thickness $\delta\sim\Ha^{-1/2}$ is used as the effective horizontal
length scale of instability in the $(x,z)$ plane.

We found that the conductivity of Hartmann walls has a significant
stabilizing effect on the flow as long as $c\Ha\lesssim30.$ Since
the instability originates in the side-wall jets with the characteristic
thickness $\delta\sim\Ha^{-1/2},$ the critical Reynolds number scales
as $\RE_{c}\sim\Ha^{1/2}\tilde{\RE}_{c}(c\Ha),$ where $\tilde{\RE}_{c}$
is a rescaled critical Reynolds number which depends mainly on $c\Ha$
and varies very little with $\Ha.$ A similar relationship holds also
for the critical wavenumber $k_{c}\sim\Ha^{1/2}\tilde{k}_{c}(c\Ha),$
where the rescaled wavenumber $\tilde{k}_{c}$ starts depend directly
on $\Ha$ if $c\Ha\lesssim10.$ This is likely due to the effect of
the core flow, which has the relative velocity $\sim(c\Ha)^{-1}$,
and thus may be no longer negligible with respect to the jet velocity
if $c\Ha\lesssim10.$ In the strong magnetic field satisfying $c\Ha\gtrsim30$,
the stabilizing effect of wall conductivity vanishes, and the asymptotic
solution $\tilde{\RE}_{c}\approx90$ and $\tilde{k}_{c}\approx0.525$
found by \citet{Priede2010} for the ideal Hunt's flow with perfectly
conducting Hartmann walls is recovered. Consequently, for the Hartmann
walls to become virtually perfectly conducting with respect to the
stability of flow, the magnetic field with $\Ha\gtrsim30/c$ is required.
Maximal destabilization of the flow, i.e., the lowest critical Reynolds
number for the given wall conductance ratio, is achieved at $\Ha\approx30/c.$
This result suggests an optimal design of liquid metal blankets when
efficient turbulent removal of heat from side walls is required. Note
that a much stronger magnetic field with $\Ha\gg c^{-2}$ is required
for the volume flux carried by the side-wall jets to fully develop
and become dominant as in the Hunt's flow with perfectly conducting
Hartmann walls. However, it is the local velocity distribution at
the side walls which determines the stability of this type of flow.
Therefore it is the relative velocity of the core flow rather than
its volume flux which is relevant for the stability of Hunt's flow. 

In conclusion, it is important to note that the instability considered
in this study corresponds to the so-called convective instability
\citep[Secs. 7.2.1--7.2.3]{Schmid2012}. In contrast to the absolute
instability, the convective one is not in general self-sustained and,
thus, may not be directly observable in the experiments without external
excitation. However, it should be observable in the direct numerical
simulation using periodicity condition in the stream-wise direction.
The absolute instability threshold, if any, is still unknown for this
type of flow. But given the relatively low local critical Reynolds
number $\tilde{\RE}_{c}\sim100,$ it is very likely to be relevant
for such MHD duct flows with the side-wall jets.

Also the physical mechanism behind in the instability itself is not
entirely clear. The low $\tilde{\RE}_{c}$ as well as the presence
of inflection points in the velocity profile imply that the instability
could be inviscid, although the latter criterion is strictly applicable
to one-dimensional inviscid flows only. The respective criteria of
two-dimensional inviscid flows are considerably more complicated \citep{Bayly1988},
and no such criterion is known for MHD flows. A complicated numerical
analysis may be required to answer this non-trivial question.

This work was supported by the Liquid Metal Technology Alliance (LIMTECH) of the Helmholtz Association. The authors are indebted to the Faculty of Engineering, Environment and Computing of Coventry University for the opportunity to use its high performance computer cluster.

\section{\label{sec:nume}Vector streamfunction-vorticity formulation}

In order to satisfy the incompressibility constraint $\vec{\nabla}\cdot\vec{v}=0$,
we introduce a vector streamfunction $\vec{\psi}$ which allows us
to seek the velocity distribution in the form $\vec{v}=\vec{\nabla}\times\vec{\psi}.$
Since $\vec{\psi}$ is determined up to a gradient of arbitrary function,
we can impose an additional constraint 
\begin{equation}
\vec{\nabla}\cdot\vec{\psi}=0,\label{eq:divpsi}
\end{equation}
which is analogous to the Coulomb gauge for the magnetic vector potential
$\vec{A}$ \citep{Jackson1998}. Similar to the incompressibility
constraint for $\vec{v},$ this gauge leaves only two independent
components of $\vec{\psi}.$ 

The pressure gradient is eliminated by applying \textit{curl} to (\ref{eq:NS}).
This yields two dimensionless equations for $\vec{\psi}$ and $\vec{\omega}$
\begin{eqnarray}
\partial_{t}\vec{\omega} & = & \vec{\nabla}^{2}\vec{\omega}-\RE\vec{g}+\Ha^{2}\vec{h},\label{eq:omeg}\\
0 & = & \vec{\nabla}^{2}\vec{\psi}+\vec{\omega},\label{eq:psi}
\end{eqnarray}
 where $\vec{g}=\vec{\nabla}\times(\vec{v}\cdot\vec{\nabla})\vec{v},$
and $\vec{h}=\vec{\nabla}\times\vec{f}$ are the \textit{curls} of
the dimensionless convective inertial and electromagnetic forces,
respectively. 

The boundary conditions for $\vec{\psi}$ and $\vec{\omega}$ were
obtained as follows. The impermeability condition applied integrally
as $\int_{s}\vec{v}\cdot\vec{ds}=\oint_{l}\vec{\psi}\cdot\vec{dl}=0$
to an arbitrary area of wall $s$ encircled by the contour $l$ yields
$\left.\psi_{\tau}\right|_{s}=0.$ This boundary condition substituted
into (\ref{eq:divpsi}) results in $\left.\partial_{n}\psi_{n}\right|_{s}=0.$
In addition, applying the no-slip condition integrally $\oint_{l}\vec{v}\cdot\vec{\mathrm{d}l}=\int_{s}\vec{\omega}\cdot\vec{\mathrm{d}s}$
we obtain $\left.\omega_{n}\right|_{s}=0.$

The base flow can conveniently be determined using the $z$-component
of the induced magnetic field $\bar{b}$ instead of the electrostatic
potential $\bar{\phi}.$ Then the governing equations for the base
flow take the form 
\begin{eqnarray}
\vec{\nabla}^{2}\bar{w}+\Ha\partial_{y}\bar{b} & = & \bar{P,}\label{eq:wbar}\\
\vec{\nabla}^{2}\bar{b}+\Ha\partial_{y}\bar{w} & = & 0,\label{eq:bbar}
\end{eqnarray}
where $\Ha=dB\sqrt{\sigma/(\rho\nu)}$ is the Hartmann number and
$\bar{b}$ is scaled by $\mu_{0}\sqrt{\sigma\rho\nu^{3}}/d.$ The
constant dimensionless axial pressure gradient $\bar{P}$ that drives
the flow is determined from the normalisation condition $\bar{w}_{\max}=1.$
The velocity satisfies the no-slip boundary condition $\bar{w}=0$
at $x=\pm1$ and $y=\pm A,$ where $A=h/d$ is the aspect ratio, which
is set equal to $1$ for the square cross-section duct considered
in this study. The boundary condition for the induced magnetic field
\citep{Shercliff1956} at the Hartmann wall is 
\begin{equation}
\bar{b}=c\partial_{n}\bar{b}.\label{bc:bbar}
\end{equation}
and $\bar{b}=0$ for the side wall.

Linear stability of the base flow $\{\bar{\vec{\psi}},\bar{\vec{\omega}},\bar{\phi}\}(x,y)$
is analysed with respect to infinitesimal disturbances in the standard
form of harmonic waves travelling along the axis of the duct \global\long\def\i{\mathrm{i}}
 
\[
\{\vec{\psi},\vec{\omega},\phi\}(\vec{r},t)=\{\bar{\vec{\psi}},\bar{\vec{\omega}},\bar{\phi}\}(x,y)+\{\hat{\vec{\psi}},\hat{\vec{\omega}},\hat{\phi}\}(x,y)e^{\lambda t+\i kz},
\]
 where $k$ is a real wavenumber and $\lambda$ is, in general, a
complex growth rate. This expression substituted into (\ref{eq:omeg},\ref{eq:psi})
results in
\begin{eqnarray}
\lambda\hat{\vec{\omega}} & = & \vec{\nabla}_{k}^{2}\hat{\vec{\omega}}-\textit{Re}\hat{\vec{g}}+\textit{Ha}^{2}\hat{\vec{h}},\label{eq:omegh}\\
0 & = & \vec{\nabla}_{k}^{2}\hat{\vec{\psi}}+\hat{\vec{\omega}},\label{eq:psih}\\
0 & = & \vec{\nabla}_{k}^{2}\hat{\phi}-\hat{\omega}_{\shortparallel},\label{eq:phih}
\end{eqnarray}
 where $\vec{\nabla}_{k}\equiv\vec{\nabla}_{\perp}+\i k\vec{e}_{z};$
$\shortparallel$ and $\perp$ respectively denote the components
along and transverse to the magnetic field in the $(x,y)$-plane.
Because of the solenoidality of $\hat{\vec{\omega}},$ we need only
the $x$- and $y$-components of (\ref{eq:omegh}), which contain
$\hat{h}_{\perp}=-\partial_{xy}\hat{\phi}-\partial_{\shortparallel}\hat{w},$
$\hat{h}_{\shortparallel}=-\partial_{\shortparallel}^{2}\hat{\phi}$
and 
\begin{eqnarray}
\hat{g}_{x} & = & k^{2}\hat{v}\bar{w}+\partial_{yy}(\hat{v}\bar{w})+\partial_{xy}(\hat{u}\bar{w})+\i2k\partial_{y}(\hat{w}\bar{w}),\label{eq:gx}\\
\hat{g}_{y} & = & -k^{2}\hat{u}\bar{w}-\partial_{xx}(\hat{u}\bar{w})-\partial_{xy}(\hat{v}\bar{w})-\i2k\partial_{x}(\hat{w}\bar{w}),\label{eq:gy}
\end{eqnarray}
 where 
\begin{eqnarray}
\hat{u} & = & \i k^{-1}(\partial_{yy}\hat{\psi}_{y}-k^{2}\hat{\psi}_{y}+\partial_{xy}\hat{\psi}_{x}),\label{eq:uh}\\
\hat{v} & = & -\i k^{-1}(\partial_{xx}\hat{\psi}_{x}-k^{2}\hat{\psi}_{x}+\partial_{xy}\hat{\psi}_{y}),\label{eq:vh}\\
\hat{w} & = & \partial_{x}\hat{\psi}_{y}-\partial_{y}\hat{\psi}_{x}.\label{eq:wh}
\end{eqnarray}
 The boundary conditions are
\begin{eqnarray}
\partial_{x}\hat{\phi}=\hat{\psi}_{y}=\partial_{x}\hat{\psi}_{x}=\partial_{x}\hat{\psi}_{y}-\partial_{y}\hat{\psi}_{x}=\hat{\omega}_{x}=0 & \mbox{ at } & x=\pm1,\label{eq:BC1}\\
\partial_{x}^{2}\hat{\phi}-k^{2}\hat{\phi}\mp c^{-1}\partial_{y}\hat{\phi}=\hat{\psi}_{x}=\partial_{y}\hat{\psi}_{y}=\partial_{x}\hat{\psi}_{y}-\partial_{y}\hat{\psi}_{x}=\hat{\omega}_{y}=0 & \mbox{ at } & y=\pm A,\label{eq:BC2}
\end{eqnarray}
where the aspect ratio $A=h/d=1$ for the square cross-section duct
considered in this study. The problem was solved using the same spectral
collocation method as in our previous study \citep{PriedeArltBuehler2015}.

Owing to the double reflection symmetry of the base flow with respect
to the $x=0$ and $y=0$ planes, small-amplitude perturbations with
different parities in $x$ and $y$ decouple from each other. This
results in four mutually independent modes, which we classify as $(o,o),$
$(o,e),$ $(e,o),$ and $(e,e)$ according to whether the $x$ and
$y$ symmetry of $\hat{\psi}_{x}$ is odd or even, respectively. Our
classification of modes corresponds to the symmetries I, II, III,
and IV used by \citet{Tatsumi1990} and \citet{Uhlmann2006} (see
table \ref{tab:mod}). The symmetry allows us to solve the linear
stability problem for each of four modes separately using only one
quadrant of the duct cross-section. A more detailed description of
the spatial structure of different instability modes can be found
in \citet{PriedeArltBuehler2015}.
\begin{table}
\begin{centering}
\begin{tabular}{rcccc}
 & I & II & III & IV\tabularnewline
$\hat{\psi}_{x},\,\hat{\omega}_{x},\,\hat{v}:$ & $(o,o)$ & $(o,e)$ & $(e,o)$ & $(e,e)$\tabularnewline
$\hat{w}:$ & $(o,e)$ & $(o,o)$ & $(e,e)$ & $(e,o)$\tabularnewline
$\hat{\psi}_{z},\,\hat{\omega}_{z}:$ & $(e,o)$ & $(e,e)$ & $(o,o)$ & $(o,e)$\tabularnewline
$\hat{\psi}_{y},\,\hat{\omega}_{y},\,\hat{u},\,\phi:$ & $(e,e)$ & $(e,o)$ & $(o,e)$ & $(o,o)$\tabularnewline
\end{tabular}
\par\end{centering}
\caption{\label{tab:mod}The $(x,y)$ parities of different variables for symmetries
I, II, III and IV; $e$ - even, $o$ - odd }
\end{table}

\bibliographystyle{jfm}
\bibliography{Duct_Hunt}

\end{document}